\documentclass[11pt,a4paper]{amsart}
\usepackage[T1]{fontenc}
\usepackage[utf8]{inputenc}
\usepackage{lmodern,microtype}
\usepackage[margin=29mm]{geometry}
\usepackage{booktabs,tabularx,array,float}
\usepackage{xurl}
\usepackage[hidelinks]{hyperref}
\newcolumntype{Y}{>{\raggedright\arraybackslash}X}
\newcommand{\US}{\_\allowbreak}
\selectfont

\title[Proof checking on AArch64]{From Proof Construction to Independent Checking: Reproducing the Lean Verification of RH and Dirichlet GRH on AArch64}
\author{Zolt\'an G\'abor Moln\'ar}
\address{Department of Algebra and Geometry, Institute of Mathematics, Faculty of Natural Sciences, Budapest University of Technology and Economics, Budapest, Hungary}
\email{molnar.zoltan.gabor@ttk.bme.hu}
\date{}
\hypersetup{pdftitle={From Proof Construction to Independent Checking: Reproducing the Lean Verification of RH and Dirichlet GRH on AArch64},pdfauthor={Zoltan Gabor Molnar}}

\begin{document}
\begin{abstract}
Constructing a formal proof and checking a completed one are distinct computational tasks. We examine their practical separation through a reproduction of the compact Lean formalization of RH and Dirichlet GRH. On an NVIDIA DGX Spark (AArch64), with execution restricted to logical CPUs 0--7, the source build finished in 48 min 44 s and the sandboxed paranoid Comparator workflow in 3 h 47 min 15 s. The five formal targets passed the Comparator's checks, and seven checker invocations accepted the exported solution. Pinned source revisions, logs, and checksums document the result: a full multi-checker verification conducted on a single workstation.
\end{abstract}
\maketitle

\section{Construction versus checking}

The resources needed to \emph{construct} a mathematical proof can differ substantially from those needed to \emph{check} an explicit proof. In intuitionistic propositional logic, Statman's theorem establishes PSPACE-completeness of provability even for the implicational fragment \cite{statman}. In the Cook--Reckhow framework, efficient verification of explicit certificates is part of the definition of a propositional proof system \cite{cookreckhow}. Large formal developments in dependent type theory bring a further practical dimension: checking elaborated proof terms involves processing extensive dependencies and performing definitional-equality computations \cite{carneiro}.

The Lean formalization of the Riemann hypothesis (RH) and of the generalized Riemann hypothesis for Dirichlet $L$-functions (GRH) gives this distinction a concrete setting. Alp\"oge's original release comprised 17,374,215 physical source lines; the compact edition by Sernau and collaborators contains 112,827 physical lines, including 76,149 nonblank, noncomment code lines \cite{community,sernau,sutherland}. While these figures concern the construction and representation of the proof, its subsequent validation is a separate computational task: the checkers process elaborated declarations and their dependencies. Here we record the resources and verdicts of one complete multi-checker verification of the compact edition on an AArch64 workstation.

\begin{table}[H]
\caption{Reported checks of the compact Lean edition.}\label{tab:prior}
\footnotesize
\renewcommand{\arraystretch}{1.02}
\begin{tabularx}{\textwidth}{@{}>{\raggedright\arraybackslash}p{0.19\textwidth}Y>{\raggedright\arraybackslash}p{0.32\textwidth}@{}}
\toprule
Report & Platform & Comparator result \\
\midrule
Sutherland \cite{sutherland} & 16-core AMD Ryzen workstation (x86-64) & Full sandboxed paranoid check accepted. \\
Sonoda \cite{sonoda} & RIKEN compute node (AArch64) & Standard Comparator check accepted. \\
Rubinstein \cite{rubinstein} & Shared Waterloo compute server & Paranoid check accepted without sandbox. \\
Present run & NVIDIA DGX Spark (AArch64); logical CPUs 0--7 eligible & Full sandboxed paranoid check accepted. \\
\bottomrule
\end{tabularx}
\end{table}

Sutherland's 16-core workstation run already establishes that full paranoid checking of the compact edition does not require a computing cluster; he also obtained a successful result on AWS hardware \cite{sutherland}. Sonoda's AArch64 run used the standard Comparator \cite{sonoda}. Rubinstein completed a paranoid run on a shared Waterloo server without sandboxing, which the cluster disallowed \cite{rubinstein}. Our run adds a sandboxed paranoid check on AArch64, with execution restricted to eight of the Spark's 20 CPU cores. The archived logs record the checker acceptances and the final Comparator verdict.

\section{Statements and verification procedure}

We used the compact \texttt{lsernau/grh-lean} revision \texttt{72fead10}, Lean 4.35.0, and Mathlib revision \texttt{56a00585} \cite{sernau,mathlib}. The Comparator configuration names five statements (Table~\ref{tab:targets}). The Dirichlet theorems quantify over complex-valued characters modulo nonzero natural numbers; ``GRH'' here does not refer to arbitrary $L$-functions. The exact propositions are defined by the pinned sources.

\begin{table}[H]
\caption{The five formal targets, in abbreviated mathematical form.}\label{tab:targets}
\footnotesize
\renewcommand{\arraystretch}{1.03}
\begin{tabularx}{\textwidth}{@{}>{\raggedright\arraybackslash}p{0.52\textwidth}Y@{}}
\toprule
Lean declaration & Mathematical assertion \\
\midrule
\texttt{riemannHypothesis} & Mathlib's RH proposition. \\
\texttt{riemannZeta\US eq\US zero\US re} & Zeta zeros, excluding negative even integers and $s=1$, have real part $1/2$. \\
\texttt{LFunction\US ne\US zero\US of\US half\US lt\US re} & Dirichlet $L$-functions do not vanish for $\operatorname{Re}(s)>1/2$, except at the principal-character pole. \\
\texttt{LFunction\US eq\US zero\US re} & Zeros in $0<\operatorname{Re}(s)<1$ lie on the critical line. \\
\texttt{LFunction\US eq\US zero\US re\US of\US isPrimitive} & Corresponding statement for primitive characters. \\
\bottomrule
\end{tabularx}
\end{table}

Comparator constructs two exported environments: a \emph{challenge} with the five statements and placeholder proofs, and a \emph{solution} with the actual proofs. The version-pinned workflow compares declaration names, universe parameters, types and relevant dependencies, then audits the axioms used by the solution before invoking several proof checkers \cite{comparator}. Its axiom whitelist consists of \texttt{propext}, \texttt{Quot.sound} and \texttt{Classical.choice}; \texttt{sorryAx} is not permitted on the solution side. The log records the checker verdicts explicitly, while the successful overall verdict establishes completion of the preceding comparison and axiom-audit stages in this implementation.

\section{An eight-core paranoid check}

The experiments were performed on an NVIDIA DGX Spark with 20 ARM CPU cores. The recovered execution record specifies affinity to logical CPUs 0--7 for both runs; no GPU computation was used. It also specifies \texttt{MemoryMax} ceilings of 48~GiB for the source build and 72~GiB for Comparator, with swap disabled. The source build reused precompiled Mathlib artifacts; Comparator used a separate checkout and a Bubblewrap sandbox, compiling additional dependencies \cite{archive}. Table~\ref{tab:runs} reports the observed whole-command durations and outcomes.

\begin{table}[H]
\caption{Spark run results and configured limits.}\label{tab:runs}
\footnotesize
\renewcommand{\arraystretch}{1.05}
\begin{tabularx}{\textwidth}{@{}Y>{\raggedright\arraybackslash}p{0.28\textwidth}>{\raggedright\arraybackslash}p{0.32\textwidth}@{}}
\toprule
 & Source build & Paranoid Comparator \\
\midrule
Time / exit code & 48 min 44 s / 0 & 3 h 47 min 15 s / 0 \\
Lake jobs & 9,843 & 9,056 challenge; 9,841 solution \\
Memory ceiling (configured) & 48 GiB & 72 GiB \\
Scope & 794 compact modules & Build, export, audits, checkers \\
Outcome & Build succeeded & \texttt{Your solution is okay!} \\
\bottomrule
\end{tabularx}
\end{table}

The log records acceptance in seven checker invocations, in this order: \texttt{nanoda}, Lean paranoid, \texttt{lean4lean}, \texttt{nanoda} again, \texttt{con-leche}, \texttt{con-ron}, and Lean default. Thus \texttt{nanoda} ran twice and the two Lean kernels are related implementations. The \texttt{lean4lean} output reports 204,580 checked declarations; \texttt{con-leche} and \texttt{con-ron} each report 204,583 accepted declarations in verified mode. These counts describe the exported dependency environment rather than the number of RH/GRH targets, which is five. The log also contains 8,625 \texttt{Built Mathlib...} build records.

The Zenodo archive \cite{archive} contains sanitized build and Comparator logs (\texttt{rh-build.log} and \texttt{comparator-retry.log}), source revision pins, reconstructed reproduction procedures, and SHA-256 manifests. Table~\ref{tab:evidence} identifies the decisive log lines. Redaction of transient identifiers preserves their numbering; all 34 files covered by the archive's internal checksum manifest passed validation. The separate evidence inventory identifies which configuration details come from recovered terminal records rather than from the logs themselves.

\begin{table}[H]
\caption{Evidence anchors: B = \texttt{rh-build.log}; C = \texttt{comparator-retry.log}.}\label{tab:evidence}
\footnotesize
\renewcommand{\arraystretch}{1.02}
\begin{tabularx}{\textwidth}{@{}>{\raggedright\arraybackslash}p{0.43\textwidth}Y@{}}
\toprule
Claim & One-based log line(s) \\
\midrule
Build success and exit status & B:1, 2797--2799 \\
Both five-target exports & C:9072, 11869 \\
Seven acceptance messages & C:11871, 11873, 11987, 11989, 11992, 11995, 11997 \\
Checker declaration totals & C:11986, 11991, 11994 \\
Final verdict and exit status & C:1, 11998--12000 \\
\bottomrule
\end{tabularx}
\end{table}

\section{Interpretation}

The Spark run shows that the complete paranoid workflow is executable on a single AArch64 workstation with execution restricted to eight logical CPUs. Its 3 h 47 min 15 s duration covers dependency compilation, proof export, statement and axiom checks, and the checker invocations. The configured memory ceilings are not measurements of peak use. The complete check thus finished on an independently operated workstation, without use of a computing cluster.

\begin{samepage}
For a small illustration of subsequent use, I also imported the cached RH theorem from the separate community repository (revision \texttt{7c4ccbd6}, Lean 4.37.0-rc1) on a Dell Latitude 7390 with about 14~GiB RAM. Its cache included 802 RH modules. A Lean file importing \texttt{Main} and proving \texttt{RiemannHypothesis} by \texttt{exact riemannHypothesis} completed in 28.319 seconds (exit 0), excluding cache preparation. This illustrates the use of an imported theorem after its compiled artifacts have been made available.
\par
\end{samepage}

\section{Conclusion}

Once a formal proof exists, checking it becomes a computational task separate from the work that produced it. Sutherland demonstrated complete paranoid checking of the compact RH/GRH formalization on a 16-core x86-64 workstation. Our AArch64 reproduction completed the same workflow with execution restricted to eight logical CPUs and provides an inspectable record of the checks. Together, these runs demonstrate that a completed large formal proof can be rigorously rechecked on independently operated workstations.

\noindent\textbf{AI assistance.} AI tools assisted workflow planning, evidence organization and drafting; the reported verdicts were produced by the executed proof checkers.

\end{document}